\documentclass[twocolumn,aps,pra,superscriptaddress]{revtex4-2}
\usepackage[utf8]{inputenc}
\usepackage[normalem]{ulem}
\usepackage{amsmath}
\usepackage{amssymb}
\usepackage{gensymb}
\usepackage{bm}
\usepackage{graphicx}
\usepackage{float}
\usepackage{subfigure}
\usepackage[dvipsnames]{xcolor}
\usepackage{placeins}
\usepackage{braket}
\usepackage[colorlinks=true, linkcolor=blue, citecolor=blue, urlcolor=blue, breaklinks=true]{hyperref}

\usepackage{bbm}

\usepackage{lipsum}
\usepackage{physics}
\usepackage{dsfont}

\definecolor{Mycolor1}{HTML}{44aa99}
\definecolor{Mycolor2}{HTML}{cc6677}
\usepackage{orcidlink}

\begin{document}

\title{Time-Fractional Schrödinger Evolution in Coupled Double Quantum Dots: Memory Effects on Quantum Resources}

\author{Abdessamie Chhieb~\!\!\orcidlink{0009-0007-0711-4747}}
\email{chhiebabdessamie@gmail.com}
\affiliation{Laboratory of Theoretical Physics, Particles, Modeling and Energies, Department of Physics, Faculty of Sciences,\\ Mohamed First University, Oujda, Morocco}

\author{Mostafa Mansour~\!\!\orcidlink{0000-0003-0821-0582}}
\email{mostafa.mansour.fsac@gmail.com}
\affiliation{Laboratory of High Energy Physics and Condensed Matter, Department of Physics,\\ Faculty of Sciences of Aïn Chock, Hassan II University,\\ P.O. Box 5366 Maarif, Casablanca 20100, Morocco}

\author{Mohamed Ouchrif~\!\!\orcidlink{0000-0002-2954-1420}}
\email{ouchrif@cern.ch}
\affiliation{Laboratory of Theoretical Physics, Particles, Modeling and Energies, Department of Physics, Faculty of Sciences,\\ Mohamed First University, Oujda, Morocco}

\begin{abstract}
Our work explore the time evolution of entanglement, local quantum uncertainty, and correlated coherence, within a system modeled by two double quantum dots. The dynamics is represented using a time-fractional Schrödinger equation, which includes memory effects in a non-Markovian regime. We vary the fractional parameter $\tau$, the tunneling amplitudes $\delta_A$ and $\delta_B$, as well as the inter-dot interaction strength $\mathcal{V}$, to investigate how these key parameters govern the generation, stabilization, and decay of quantum resources within the system.
The obtained results reveal that, for both initial states, fractional dynamics with a low $\tau$ rapidly generates entanglement expecting maximal values $\mathcal{LN}\approx 1$ and non-classical correlations quantified by local quantum uncertainty. Conversely, higher values of $\tau$ lead to slower entanglement but memory effects allow quantum resources to remain significant for a longer time, with the negativity remaining above ($\approx 0.6$). We also find that higher interaction frequencies $\mathcal{V}$ accelerate correlations and stabilize coherence, while a strong tunneling asymmetry degrades entanglement and coherence despite the initial benefits of increasing quantum resources.
 
\end{abstract}



\maketitle


\section{Introduction}
Double quantum dots (DQDs), specifically in the form of a quantum dot molecule, are currently the center of quantum mechanical phenomena research such as entanglement and electronic correlation~\cite{bennett2014quantum, cruz2022quantum, zhang2019powerful,bennett1993teleporting, bouwmeester1997experimental, keet2010quantum}. With their ability for spin and charge confinement and control at the nanoscale level, DQDs are also leading candidates for implementation in quantum computation and communication. these quantum advantages are countered by environmental fluctuations—such as charge noise and phonon coupling—which induce decoherence and degradation of quantum correlations~\cite{de2024quantum, keller2014emergent, ferreira2023thermal}. However, DQDs have enabled auspicious functionalities such as coherent control of the spins, quantum memory, and teleportation protocols~\cite{mansour2020quantum, majek2021majorana, zhang2021anisotropic, ginzel2020spin, troiani2000exploiting}, which are supporting the establishment of scalable quantum architectures on semiconductor platforms~\cite{flagan1991fabrication, troiani2000exploiting}.

Simultaneously, The rise of fractional quantum mechanics has yielded new theoretical means of simulation of complex quantum systems. The fractional Schrödinger equation (FSE), originally formulated by Laskin~\cite{laskin2002fractional}, involves fractional derivatives to model anomalous diffusion and memory phenomena~\cite{Hilfer2000}. In particular, its time-fractional extension (TFSE)~\cite{naber2004time, Tarasov2011} provides one with the capability of investigating non-Markovian and dissipative quantum dynamics that deviate from pure Hamiltonian evolution. Ajouter has been used to describe physical processes, nonlinear instability, and wave packet collapse~\cite{aceves2022spatio} and has been used in, say, optical signal processing~\cite{liu2023experimental} and fractional quantum systems such as the Bohr atom and nonlocal oscillators~\cite{laskin2002fractional}. The TFSE provides a powerful paradigm to simulate open quantum systems where memory effects and long-range correlations are intrinsic features~\cite{ref4, chhieb2024metrological, chhieb2025fractional, el2022dynamics}, such as those ubiquitously observed in spin chains and topological condensed matter systems~\cite{ikeda2020nonlocal}. We want to emphasize here that TFSE is not aimed at replacing Contrary to that, It not describe the intrinsic unitary evolution of an isolated quantum system, but instead it represents a phenomenological effective framework that describes the attenuated dynamics of open quantum systems, especially in regimes where memory-dependent effects become important and the conventional Markovian master equations lose their validity~\cite{breuer2002theory, Hilfer2000}.
Unlike the usual, based on a first-order time derivative entailing a purely unitary and exponential (Markovian) evolution, the fractional time derivative introduced in the TFSE will allow a natural description of non-Markovian behaviors and long-term memory effects typical of complex quantum environments.

The use of Caputo fractional derivative is especially welcome since it is now possible to implement physically meaningful initial conditions on the wave function without any problem \cite{Hilfer2000, podlubny1998fractional}. Moreover, the resulting Solutions naturally express themselves in terms of the Mittag-Leffler functions, characterizing nonexponential relaxation together with a dynamics of power-law type, as consistently reported in several solid-state systems and superconducting devices featuring anomalous decoherence effects \cite{naber2004time, Tarasov2011, liu2023experimental}. From this viewpoint, the TFSE should then be regarded as a minimalist and straightforward encoding of reduced dynamics that are non-Markovian rather than an afterthought to the quantum mechanics postulates. The fractional parameter therefore functions as an perfect measure of the strength of memory in the interaction between subsystems, in agreement with recent experiment and theory works on non-Markovian aspects of superconducting and other solid-state devices \cite{chhieb2025time, abdessamie2025non, banouni2025thermal}. Non-classical correlation study is one of the central focus areas of modern quantum theory.

Under its scope, Local Quantum Uncertainty (LQU) has been a prominent measure of quantum correlations \cite{Girolami2013}, quantifying the minimum local measurement uncertainty and identifying discord-like quantum features that might survive without entanglement \cite{Henderson2001,Ollivier2001}.

On the other hand, Logarithmic Negativity ($\mathcal{LN}$) remains one of the most popular entanglement measures, appreciated for its computational simplicity and easy physical interpretation \cite{vidal2002computable, plenio2005logarithmic}. It estimates entanglement in bipartite systems through the spectral properties of the partially transposed density matrix, which is an efficient and robust tool for detecting and quantifying entanglement. Such nonclassical resources form the basis of secure communication protocols and distributed quantum computing~\cite{cavalcanti2009experimental, branciard2012one, popescu2014nonlocality}.Entanglement itself is one of the basics of quantum information science~\cite{horodecki2009quantum}, with robust quantifiers such as concurrence~\cite{Wootters1998}, negativity~\cite{Vidal2002}, entanglement entropy, and resource-theoretic frameworks~\cite{Horodecki1996, Plenio2005} as the basis for theoretical and experimental research. Research in the field increasingly considers the time evolution of such correlations, especially for open systems prone to decoherence. In this case, variation between Markovian and non-Markovian regimes will be a crucial factor in quenching or transiently reviving quantum coherence and correlations~\cite{costa2016generalized, bukbech2023quantum, Chouiba2025}. Double quantum dots (DQDs) are significant for quantum information processing, but the time-fractional dynamics of quantum resources is still not adequately exploited. Since realistic nanoscale environments naturally exhibit memory effects and deviations from purely Markovian evolution, the (TFSE) provides a powerful and flexible Structure for modeling such systems~\cite{Tarasov2011, ikeda2020nonlocal}.

We fill this gap here by examining the dynamics of quantum resources in a DQD system modeled by the TFSE. Specifically, we study the behavior of logarithmic negativity (LN) as an entanglement measure, local quantum uncertainty (LQU) as a quantifier for quantum correlations, and correlated coherence $(C_{cc})$ as a coherence resource, under various asymmetries and initial conditions. This study sheds light on the contribution of non-Markovian memory effects in solid-state quantum technologies~\cite{breuer2016colloquium, laine2010measure} and suggests strategies for the preservation of quantum resources and engineering more fault-tolerant, robust quantum technologies~\cite{bylicka2014non, chhieb2025fractional}. The article is structured as follows. Section~\ref{sec2} is left to present the quantifiers to characterize quantum resources.

Section~\ref{sec3} initially provides a thorough introduction to the double quantum dot (DQD) system and solution of the time-fractional Schrödinger equation. The key outcomes are investigated step by step in Section~\ref{sec4}, with particular emphasis on the evolutionary dynamics of $C_{cc}$, $C_{cc}$, and $C_{cc}$. Finally, Section~\ref{sec5} gives a brief overview of the key findings and suggests avenues for future research.
\section{Quantum resources }
\label{sec2}
Here, we introduce logarithmic negativity (LN), local quantum uncertainty (LQU), and correlated coherence ($C_{cc}$) and provide a unified perspective of quantum resource dynamics under time-fractional evolution.
	\subsection{Logarithmic Negativity}
 For bipartite quantum system represented by the density operator $\varrho$, Logarithmic negativity (LN) is defined \cite{negat,neg2}:
\begin{equation}
\label{LN1}
\mathcal{E}(\varrho) = \log_2 \left\| \varrho^{T_A} \right\|_1,
\end{equation}
where $\varrho^{T_A}$ denotes the partial transpose of the density matrix $\varrho$ with respect to subsystem~$A$.
The trace norm of an operator $O$ is defined as
\begin{equation}
\left\| O \right\|_1 = \mathrm{Tr}\!\left( \sqrt{O^{\dagger} O} \right),
\end{equation}
which is equal to the sum of singular values of $O$.
Based on this definition, the negativity can be introduced as~\cite{neg2}:
\begin{equation}
\label{neg}
\mathcal{N}(\varrho) = \frac{ \left\| \varrho^{T_A} \right\|_1 - 1 }{2}. 
\end{equation}
The negativity can also be exoressed in terms of the eigenvalues $\lambda_i$ of the partial transpose $\varrho^{T_A}$:

\begin{equation}
\label{neg2}
\mathcal{N}(\varrho) = \sum_i \max(0, -\lambda_i) = \sum_i \frac{ | \lambda_i | - \lambda_i }{2}.
\end{equation}
It then follows that the logarithmic negativity can be written as:
\begin{equation}
\label{LN2}
\mathcal{LN} = \log_2 \left( 2 \mathcal{N}(\varrho) + 1 \right).
\end{equation}
\subsection{Local Quantum Uncertainty}
 LQU ($\mathcal{L}_q$) is defined as a discord-like quantifier for characterizing non-classical correlations. It can be given for a bipartie density matrix $\rho_{AB}$, utilizing the following \cite{Girolami2013}:
\begin{equation}
\label{eq:LQU_definition}
\mathcal{LQU}= \min_{M_A} I(\varrho, M_A \otimes \mathbb{I}_B).
\end{equation}
where, the Wigner-Yanase skew information is expressed by:
\begin{equation}
\label{eq:skew_information}
I(\rho_{AB}, M) = -\frac{1}{2} \mathrm{Tr}\left(\left[\varrho^{1/2}, M\right]^2\right).
\end{equation}
For bipartite systems of size $2 \otimes d$ (with subsystem $A$ a qubit), the LQU has a closed-form expression \cite{9a}:
\begin{equation}
\label{eq:LQU_qubit}
\mathcal{U}(\varrho) = 1 - \lambda_{\max}(\mathbf{W}),
\end{equation}
where $\lambda_{\max}(\mathbf{W})$ denotes the largest eigenvalue of the real symmetric $3 \times 3$ matrix $\mathbf{W}$.
The elements of this matrix are defined as
\begin{equation}
\label{eq:W_matrix_elements}
W_{ij} = \mathrm{Tr}\!\left[
\sqrt{\varrho}\, (\sigma_i \otimes \mathbb{I}_B)\,
\sqrt{\varrho}\, (\sigma_j \otimes \mathbb{I}_B)
\right],
\quad i, j \in \{x, y, z\}.
\end{equation}

Here, $\sigma_i$ are Pauli operators on subsystem $A$ and $\mathbb{I}_B$ is the identity operator on subsystem $B$.
\subsection{Correlated Coherence}
Quantum coherence is a property of the quantum system inherent to it due to superposition principles and manifesting itself through off-diagonal elements in density matrix formalism. The basis-relative resource may be measured by the $l_1$-norm measure \cite{baumgratz2014quantifying}, providing computational ease as well as theoretical stability. For a quantum state described by the density operator $\rho$, the coherence measure is as follows:
\begin{equation}
    \mathcal{C}_{l_1}(\rho) = \sum_{i \neq j} |\rho_{ij}|,
\end{equation}
An alternative formulation expresses this measure as the minimum distance to the set of incoherent states $\mathcal{I}$:
\begin{equation}
    \mathcal{C}_{l_1}(\rho) = \min_{\delta \in \mathcal{I}} \|\rho - \delta\|_1.
\end{equation}
In a $d$-dimensional Hilbert space, the coherence measure satisfies the bound $\mathcal{C}_{l_1}(\rho) \leq d-1$. To specifically address coherence arising from inter-subsystem correlations, the correlated coherence measure is introduced. For a bipartite quantum state $\rho$, this quantity is defined as:
\begin{equation}
\mathcal{C}_{cc}(\rho) = \mathcal{C}_{l_1}(\rho) - \mathcal{C}_{l_1}(\rho_A) - \mathcal{C}_{l_1}(\rho_B),
\end{equation}
where $\rho_{A} = \mathrm{Tr}_B \rho$ and $\rho_{B} = \mathrm{Tr}_A \rho$ denote the reduced density matrices of the respective subsystems.

\section{Time-Fractional Evolution of a Two-Coupled Double Quantum Dot System}
\label{sec3}

\begin{figure}[H]
	\centering
	\includegraphics[width= 0.5\textwidth]{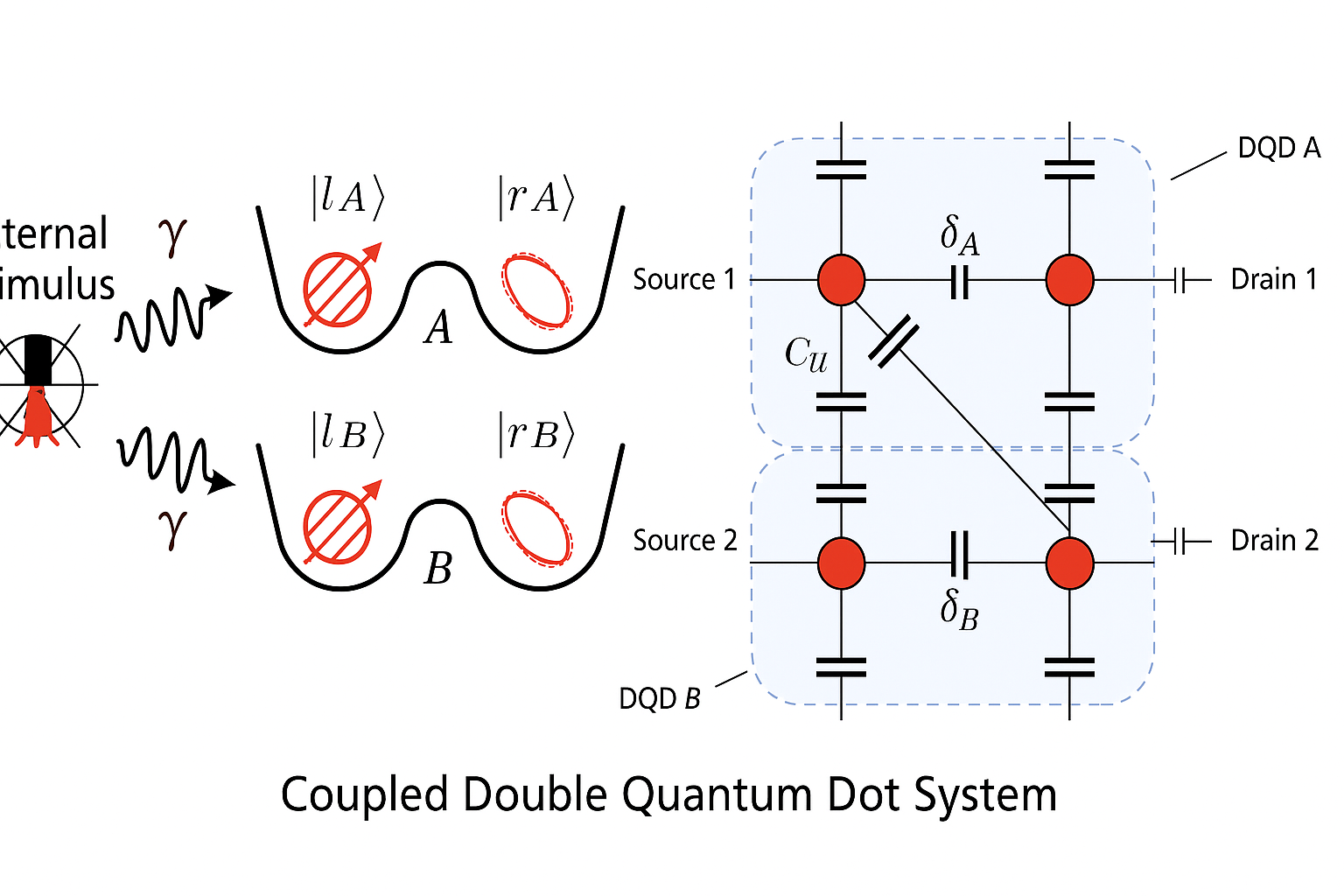}
	\caption{Schematic of two capacitively coupled double quantum dots (DQDs) and the respective equivalent circuit. The capacitive components limit electron transfer between the DQDs, while tunnel junctions enforce the Coulomb blockade regime, taking one electron at a time.}
	\label{figure1}
\end{figure}

The considered physical system is composed of two capacitively coupled double quantum dots (DQDs). Two quantum wells in each DQD can be occupied by a single electron, either in the left dot (\(\ket{\downarrow}\)) or the right dot (\(\ket{\uparrow}\)). This is a promising platform for the realization of solid-state qubits~\cite{Loss1998_QuantumComp,Hanson2007_SpinQD}, The schematic of the coupled double quantum dots is shown in Figure~\ref{figure1}.

An external perturbation in the form of an electromagnetic wave \(\gamma\) is applied to the system. It induces quantum transitions between the two electronic states depending on the amount of absorbed energy. The dynamics can be modulated by controlling the gate voltages \(V_{gA}\) and \(V_{gB}\), which adjust the potential of each quantum dot via Schottky electrodes.

In the absence of external excitation, the Hamiltonian describing the coupled DQD system is given by~\cite{ elghaayda2022local}:
\begin{equation}
H_0 = \delta_A \sigma_x^A + \delta_B \sigma_x^B + \mathcal{V} \sigma_z^A \sigma_z^B,
\end{equation}
where \(\delta_A\) and \(\delta_B\) are the tunneling amplitudes within each DQD, and \(\mathcal{V}\) represents the Coulomb interaction between the electrons in the two subsystems.
\begin{equation}
H_0 =
\begin{pmatrix}
\mathcal{V} & \delta_B & \delta_A & 0 \\
\delta_B & -\mathcal{V} & 0 & \delta_A \\
\delta_A & 0 & -\mathcal{V} & \delta_B \\
0 & \delta_A & \delta_B & \mathcal{V} 
\label{MH}
\end{pmatrix}
\end{equation}
\noindent where \(\delta_i\) is the tunneling coupling strength in each coupled quantum dot, and \(\mathcal{V}\) represents the Coulomb interaction between the electrons. To proceed, we must simplify the matrix (Eq.$\ref{MH}$) in a $X$-structured form, allowing us to use existing analytical expressions for the measure under consideration.
For achieving the matrix $X$-form ($\mathbf{H}_{X}$), we apply the double Hadamard transformation ($\bar{\mathbb{H}}$) and its conjugate transpose ($\bar{\mathbb{H}}^\dagger$) to the matrix $\mathbf{H}$, as described in \cite{chlih2024nonclassicality} $ H_{X} =\bar{\mathbb{H}}.\mathbf{H}.\bar{\mathbb{H}}^\dagger \,
     $:
     \begin{equation}\label{MX}
    H_{X} =\begin{pmatrix}
\delta_B + \delta_A & 0 & 0 & \mathcal{V}  \\
0 & \delta_A - \delta_B & \mathcal{V} & 0 \\
0 & \mathcal{V} & \delta_B - \delta_A & 0 \\
\mathcal{V}  & 0 & 0 & -\delta_B - \delta_A
\end{pmatrix}
\end{equation}
Here $\bar{\mathbb{H}}=\mathbb{H}\otimes\mathbb{H}$, and its conjugate transpose denoted as $\bar{\mathbb{H}}^{\dagger}$  with $\mathbb{H}=\frac{1}{\sqrt{2}}(\mathbf{\sigma_{x}+\sigma_{z}})$ is the Hadamard operator. the matrix $\mathbf{H}_X$
yields eigenvalues $\mathbf{\xi_j}$ and eigenstates $\ket{\phi_{j}}$ ($j=1,2,3,4$), which can be expressed in the following form:                                                                                                                                                       
\begin{align}
\label{eigenstates1} 
    \xi_1 &=-\sqrt{v^2 + \delta_A^2 - 2 \delta_A \delta_B + \delta_B^2}, \quad 
    \ket{\phi_{1}} = a \Big[ \alpha_{-}\ket{01} + \ket{10} \Big], \\
    \label{eigenstates2}
    \xi_2 &= \sqrt{v^2 + \delta_A^2 - 2 \delta_A \delta_B + \delta_B^2}, \quad 
    \ket{\phi_{2}} = b \Big[ \alpha_{+}\ket{01} + \ket{10} \Big], \\
    \label{eigenstates3}
    \xi_3 &= -\sqrt{v^2 + \delta_A^2 + 2 \delta_A \delta_B + \delta_B^2}, \quad 
    \ket{\phi_{3}} = c \Big[ \beta_{-}\ket{00} + \ket{11} \Big], \\
    \label{eigenstates4}
    \xi_4 &= \sqrt{v^2 + \delta_A^2 + 2 \delta_A \delta_B + \delta_B^2 }, \quad 
    \ket{\phi_{4}} = d\Big[ \beta_{+}\ket{00} + \ket{11} \Big],
\end{align}
where  
\begin{align*}
\alpha_{\mp} = \frac{\delta_A \mp \sqrt{v^2 + (\delta_A - \delta_B)^2} - \delta_B}{v},\\ \gamma_{\mp} = \frac{\delta_A + \delta_B \mp \sqrt{v^2 + \delta_A^2 + 2\delta_A \delta_B + \delta_B^2 }}{v},\\ 
a_{(b)} = \frac{1}{\sqrt{1 + \left| \frac{\mp\delta_A + \sqrt{v^2 + (\delta_A - \delta_B)^2} \pm \delta_B}{v} \right|^2}},\\
c_{(d)} = \frac{1}{\sqrt{1 + \left| \frac{\delta_A + \delta_B \mp \sqrt{v^2 + \delta_A^2 + 2\delta_A \delta_B + \delta_B^2 }}{v} \right|^2}},
\end{align*} 

To study the fractional-time dynamics of a two-spin system interacting via dipole-dipole coupling, we adopt the framework based on the time-fractional Schrödinger equation (TFSE), as originally introduced by Naber~\cite{naber2004time}. In this formalism, the standard time derivative of first order is replaced by the Caputo-type fractional derivative \(\mathbb{D}_t^\tau\), where the fractional order \(\tau\) satisfies \(0 < \tau \leq 1\). 
\begin{equation}
\label{FTSE}
(i)^\tau \hbar_\tau \mathbb{D}_t^\tau \, |\Psi(t, \tau)\rangle = \hat{H}_{\text{int}} |\Psi(t, \tau)\rangle,
\end{equation}
where (where $(\hbar_\tau)$ represents a generalized constant analogous to the Planck constant, and $\hat{H}_{\text{int}}$ denotes the interaction Hamiltonian. To simplify the calculations, we impose the condition $\hbar_\tau \mathbb{D}_t^\tau = 1$. Although this solution is not unique and other forms have been proposed in recent works~\cite{zu2025time, huang2025effects}, for the sake of clarity and consistency, we restrict our analysis to this type of solution. The general solution of Eq.~(\ref{FTSE}), derived in Appendix~A, can be expressed as reported in~\cite{odibat2010analytic}:
\begin{equation} 
\label{sol_general}
|\Psi(t, \tau)\rangle = \sum_{j=1}^{n} \mathcal{C}_j \, E_\tau(\xi_j t^\tau) \, |\phi_j\rangle,
\end{equation} 
where \(\lambda_j\) and \(|\phi_j\rangle\) represent the eigenvalues and eigenstates of \(\hat{H}_{\text{int}}\), and \(\mathcal{C}_j\) are arbitrary complex coefficients. The function \(E_\tau(z)\) denotes the one-parameter Mittag-Leffler function, defined as: $E_\tau(z) = \sum_{k=0}^{\infty} \frac{z^k}{\Gamma(\tau k + 1)}$,
where \(\Gamma(\tau k + 1)\) is the standard Gamma function~\cite{shukla2007generalization}. For a two-qubit system, the time-evolved state becomes:
\begin{equation}
\label{state_evolved}
|\Psi(t, \tau)\rangle = \sum_{j=1}^{4} \mathcal{C}_j \, E_\tau(\xi_j t^\tau) \, |\phi_j\rangle.
\end{equation}
We have proposed that the system is initially prepared in the pure quantum state:
\begin{equation}
\label{init_state}
|\Psi(0)\rangle = p |00\rangle + \sqrt{1 - p^2} \, |11\rangle,
\end{equation}
where \(0 \leq p \leq 1\) controls the degree of purity and population imbalance of the state, which in turn affects the quantum correlations and coherence. 

This initial state can also be decomposed in the eigenbasis of the Hamiltonian:
\begin{equation}
\label{init_decomposed}
|\Psi(0)\rangle = \sum_{j=1}^{4} \mathcal{C}_j |\phi_j\rangle,
\end{equation}
from which the coefficients \(\mathcal{C}_j\) can be extracted. Explicitly, they are given by:
\begin{equation}
\label{coeffs}
\begin{aligned}
\mathcal{C}_1 &= 0, \quad \mathcal{C}_2 = 0,   \quad \mathcal{C}_3 = \frac{p - \gamma_+ \sqrt{1 - p^2}}{a (\gamma_- - \gamma_+)}, \\
\mathcal{C}_4 &= \frac{\gamma_- \sqrt{1 - p^2} - p}{b (\gamma_- - \gamma_+)}.
\end{aligned}
\end{equation}

where \(a\), \(b\), \(\gamma_+\), and \(\gamma_-\) are constants related to the eigenstructure of \(\hat{H}_{\text{int}}\). 

Substituting these coefficients into Eq.~(\ref{state_evolved}) gives:
\begin{equation}
\label{Psi_final_reform}
\begin{aligned}
|\Psi(t, \tau)\rangle &= \chi_1(t,\tau)\, |00\rangle 
+ \chi_2(t,\tau)\, |01\rangle 
+ \chi_3(t,\tau)\, |10\rangle \\
&\quad + \chi_4(t,\tau)\, |11\rangle.
\end{aligned}
\end{equation}

with components:
\begin{align}
\chi_1(t,\tau) &= \mathcal{C}_3 \, E_\tau(\xi_3 t^\tau) \cdot a \cdot \gamma_- + \mathcal{C}_4 \, E_\tau(\xi_4 t^\tau) \cdot b \cdot \gamma_+, \\
\chi_2(t,\tau) &= 0, \\
\chi_3(t,\tau) &= 0, \\
\chi_4(t,\tau) &= \mathcal{C}_3 \, E_\tau(\xi_3 t^\tau) \cdot a + \mathcal{C}_4 \, E_\tau(\xi_4 t^\tau) \cdot b.
\end{align}

The corresponding density matrix \(\rho(t, \tau)\) is:
\begin{equation}
\label{density_matrix}
\rho(t, \tau) = \frac{1}{\mathcal{N}} \begin{pmatrix}
|\chi_1|^2 & 0 & 0 & \chi_1 \chi_4^* \\
0 & 0 & 0 & 0 \\
0 & 0 & 0 & 0 \\
\chi_4 \chi_1^* & 0 & 0 & |\chi_4|^2
\end{pmatrix},
\end{equation}
where the normalization constant is: $
\mathcal{N} = |\chi_1(t,\tau)|^2 + |\chi_4(t,\tau)|^2.$
\section{Results and Discussion\label{sec4}}  

This study examines the dynamical evolution of ogarithmic negativity ($\mathcal{LN}$), $\mathcal{LQU}$, and correlated coherence ($\mathcal{C}_{cc}$), in a system of two coupled double quantum dots (DQDs). The time-evolution is defined by a time-fractional treatment through the density matrix $\rho(\tau,t)$ obtained from the time-fractional Schrödinger equation (TFSE). Our study focuses on the role played by a host of parameters, such as the fractional order $\tau$, tunneling coupling strengths $\delta_A$ and $\delta_B$, and frequency parameter $u$, in the evolution of these quantum correlations.
The fractional order $\tau$ acts like a powerful tuning knob for quantum dynamics, as can be directly seen from the comparisons made in Figure~\ref{figure2}. While for $\tau = 1$ standard, memoryless (Markovian) evolution is precisely recovered, the reduction to $\tau < 1$ turns on strong memory effects. Those show up as an enhanced production of quantum correlations ($\mathcal{LN}$, $\mathcal{LQU}$, $\mathcal{C}_{cc}$) both for separable ($p=0$) and partially entangled ($p=0.5$) initial states. The universality of the long-time stationary state for all $\tau$ emphasizes the following key result: this parameter has an impact only on the transient path of the system while its long-time limit remains unchanged.
\begin{widetext}
	\begin{minipage}{\linewidth}
		\begin{figure}[H]
			\centering
		\subfigure[]{\label{fig1copa}\includegraphics[scale=0.35]{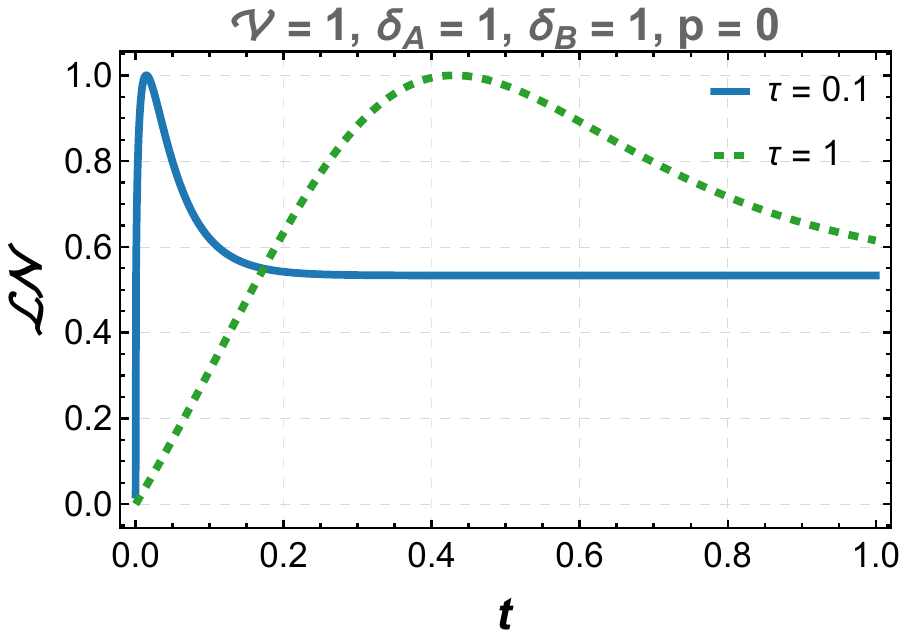}}
		\subfigure[]{\label{fig1copb}\includegraphics[scale=0.35]{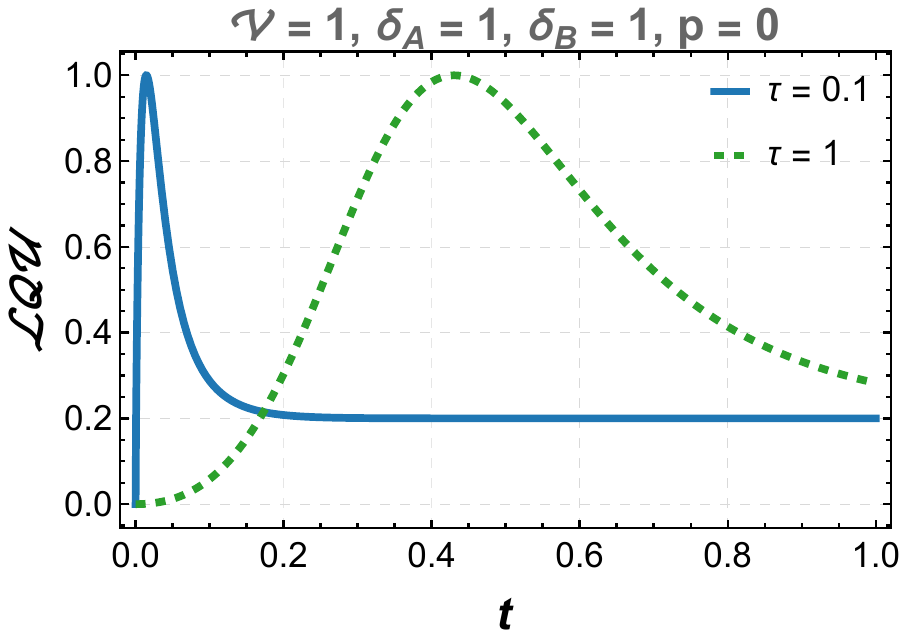}}
		\subfigure[]{\label{fig1copc}\includegraphics[scale=0.35]{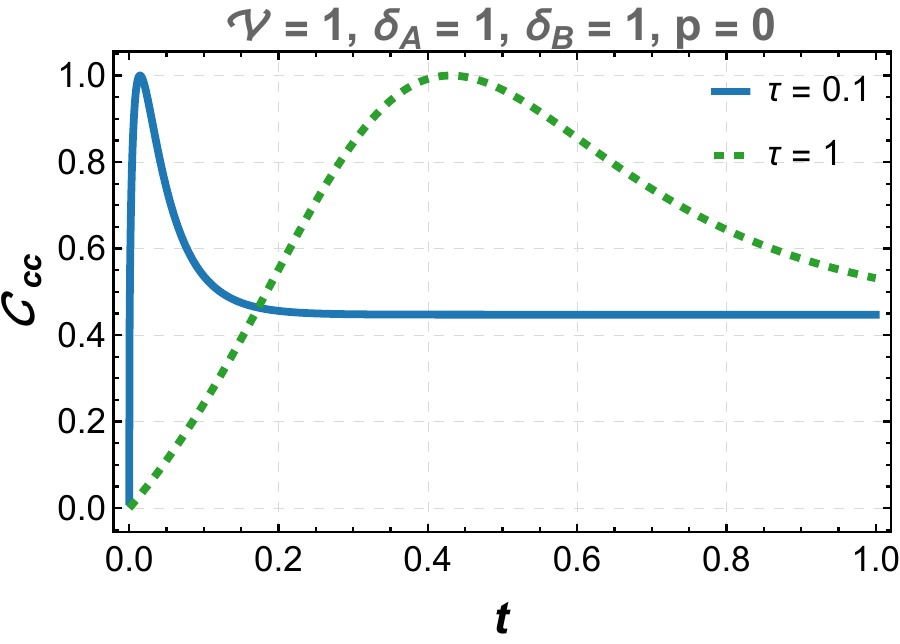}}\\
       \subfigure[]{\label{fig1copd}\includegraphics[scale=0.35]{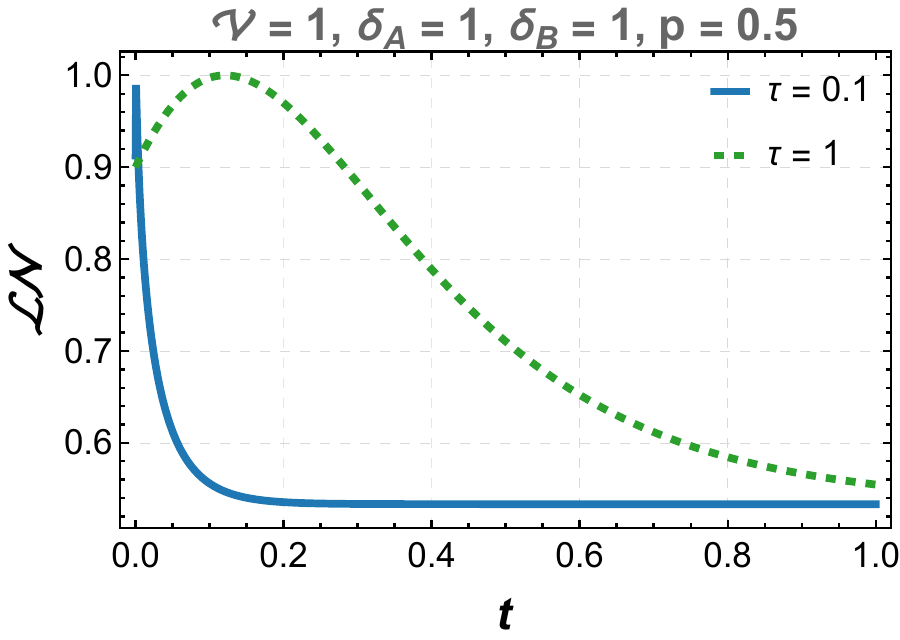}}
		\subfigure[]{\label{fig1cope}\includegraphics[scale=0.35]{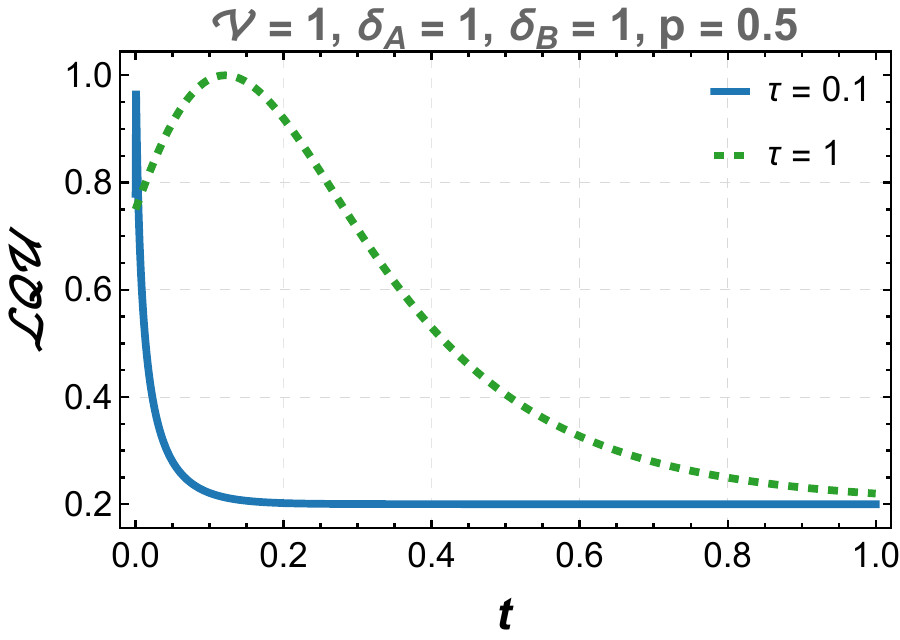}}
		\subfigure[]{\label{fig1copf}\includegraphics[scale=0.35]{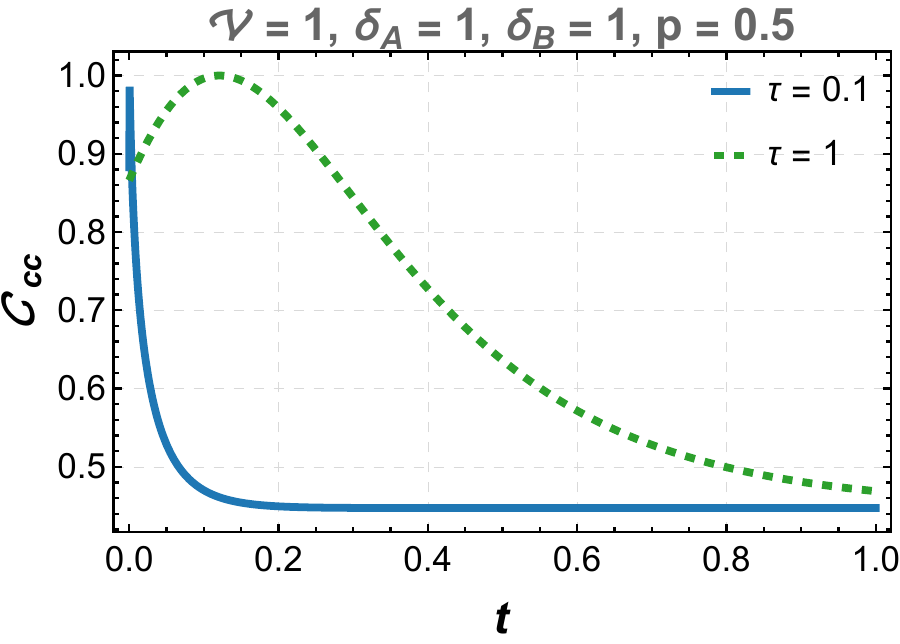}}

   \caption{Comparative plots showing the time evolution of
$\mathcal{LN}$~(\subref{fig1copa}--\subref{fig1copd}),
$\mathcal{LQU}$~(\subref{fig1copb}--\subref{fig1cope}), and
correlated coherence~(\subref{fig1copc}--\subref{fig1copf})
as functions of time $t$, for two values of the fractional parameter:
$\tau = 0.1$ (fractional dynamics) and $\tau = 1$ (standard Schrödinger evolution).
Results are obtained by taking $\mathcal{V} = 1$, $\delta_A = 1$, and $\delta_B = 1$,
with p = 0 (top panel) and p = 0.5 (bottom panel).}
\label{figure2}

\end{figure}
\end{minipage}
\end{widetext}
\begin{widetext}
	\begin{minipage}{\linewidth}
		\begin{figure}[H]
			\centering
		\subfigure[]{\label{fig1a}\includegraphics[scale=0.40]{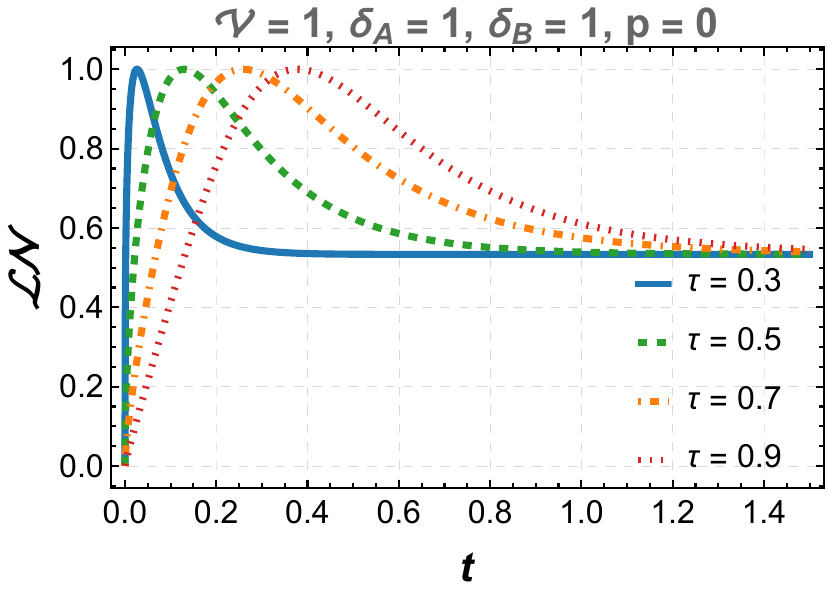}}
		\subfigure[]{\label{fig1b}\includegraphics[scale=0.40]{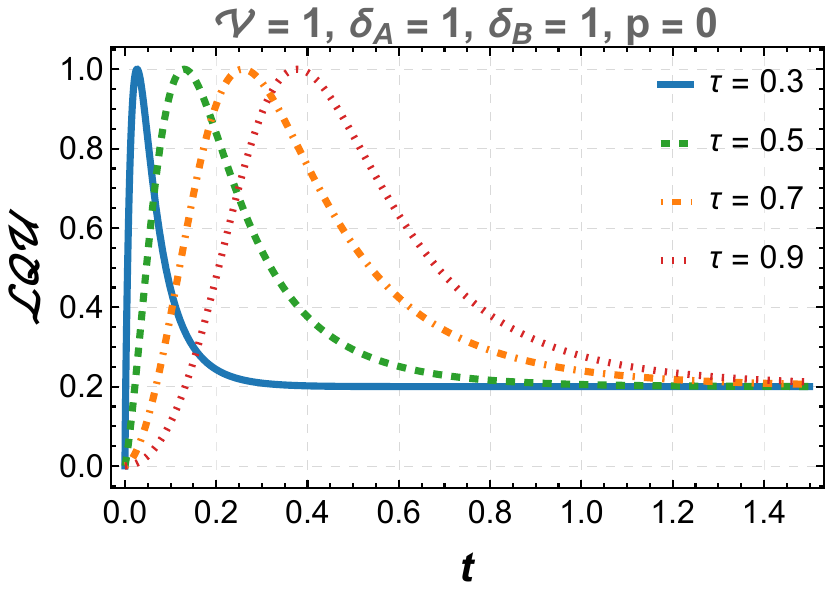}}
		\subfigure[]{\label{fig1c}\includegraphics[scale=0.40]{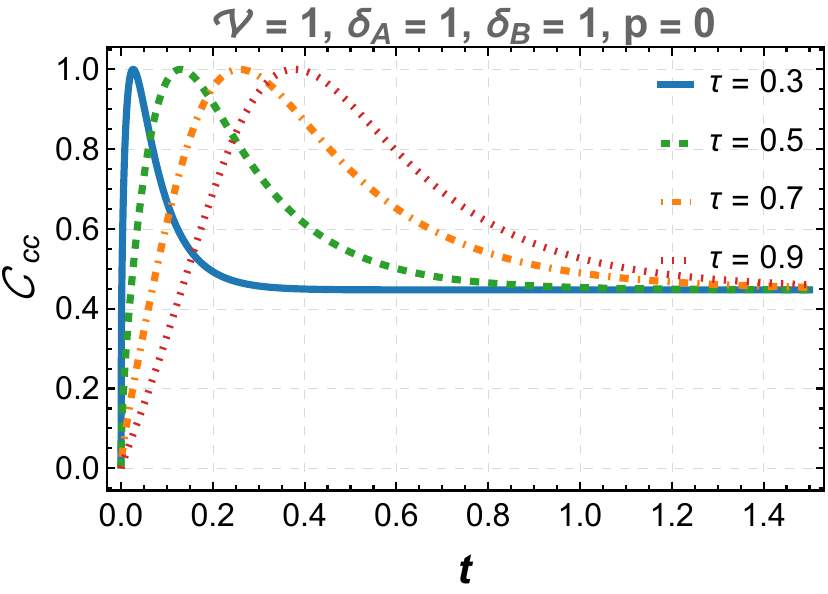}}\\
\subfigure[]{\label{fig1d}\includegraphics[scale=0.40]{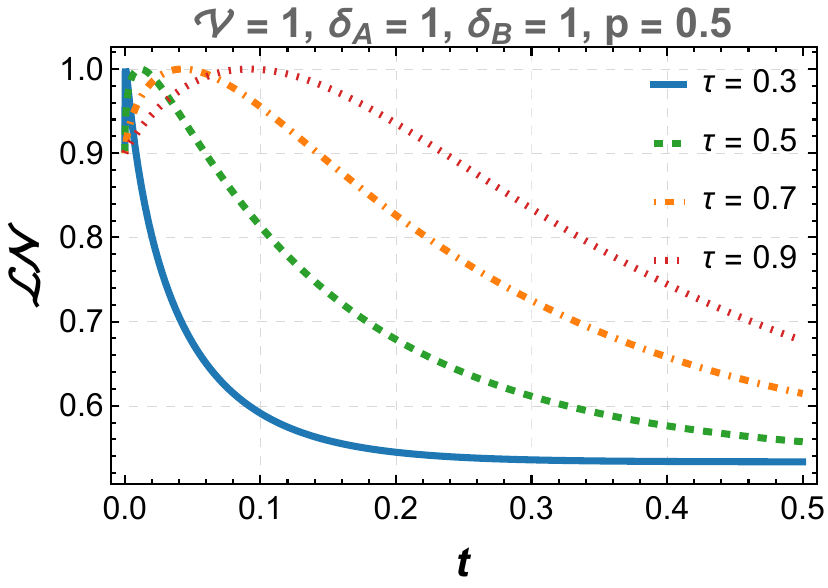}}
\subfigure[]{\label{fig1e}\includegraphics[scale=0.40]{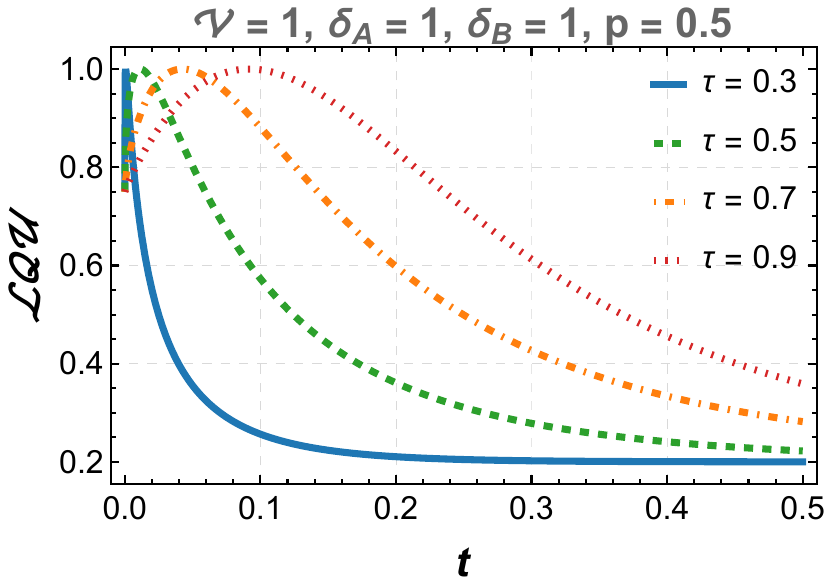}}
\subfigure[]{\label{fig1f}\includegraphics[scale=0.32]{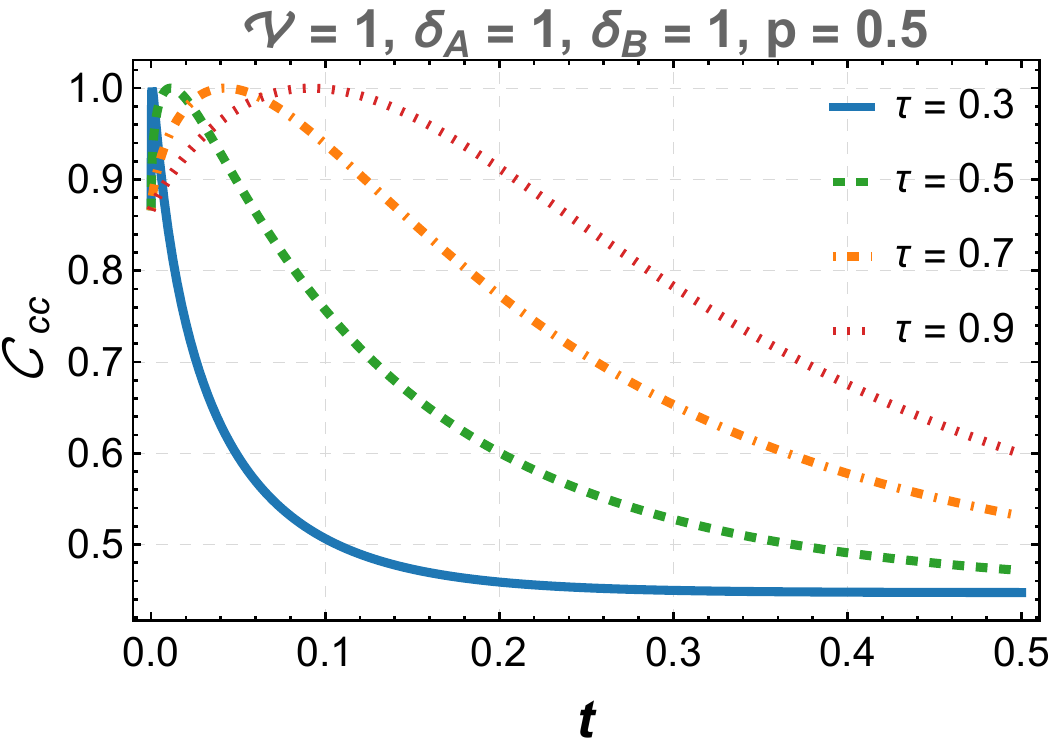}}

\caption{Plots illustrating $\mathcal{LN}$ (\subref{fig1a}--\subref{fig1d}),  
$\mathcal{LQU}$ (\subref{fig1b}--\subref{fig1e}), and correlated coherence (\subref{fig1c}--\subref{fig1f}) as functions of \(t\) for various values of the fractional parameter \(\tau\), with  
\(\mathcal{V} = 1\), \(\delta_A = 1\), \(\delta_B = 1\), for \(p=0\) \textit{(top panel)} and \(p=0.5\) \textit{(bottom panel)}.}
\label{figure3}

\end{figure}
\end{minipage}
\end{widetext}
Figure~\ref{figure3} illustrates the dynamics of quantum correlations in a system of two coupled double quantum dots (DQDs) under the influence of fractional dynamics. Three fundamental quantum resources — logarithmic negativity ($\mathcal{LN}$), local quantum uncertainty ($\mathcal{LQU}$), and correlated coherence ($\mathcal{C}_{cc}$) — are analyzed as a function of time for different values of the fractional parameter $\tau$ and for two distinct initial states ($p=0$ and $p=0.5$). For the separable initial state ($|11\rangle$), entanglement emerges rapidly ($t>0$) due to the inter-qubit coupling, confirming that entanglement can be dynamically generated from product states under the action of interactions. For $\tau=0.3$, the dynamics is close to the standard unitary evolution, leading to a fast growth of entanglement ($\mathcal{LN}\approx 1$). In contrast, for $\tau=0.9$, memory effects become dominant: the environment acts as a dissipative reservoir that limits the maximum amplitude ($\mathcal{LN}\approx 0.4$), resulting in a more dissipative evolution~\cite{tarasov2011fractional}. For the partially entangled state ($p=0.5$), correlations are initially high. Their evolution is governed by the competition between reinforcement due to memory effects and cumulative dissipation. When $\tau$ is large, memory counteracts information loss, prolonging the lifetime of entanglement ($\mathcal{LN}\approx 0.69$ still at $t\approx 0.5$). Conversely, for small $\tau$, the strongly non-Markovian regime induces faster oscillations but also an accelerated decay of correlations. These results reveal the crucial role of the fractional parameter $\tau$ as a physical control of the degree of non-Markovianity. For $\tau \ll 1$, the system behaves similarly to standard Schrödinger evolution, promoting a rapid generation of quantum resources. In the opposite regime $\tau \rightarrow 1$, memory effects enhance dissipation, reducing the amplitude and lifetime of correlations. This behavior is consistent with previous findings showing that non-Markovian environments may preserve entanglement~\cite{b14}, whereas strongly dissipative reservoirs accelerate its degradation. Fractional dynamics, recently introduced as a unified framework to describe memory-induced dissipation in open systems, thus appears as a promising approach~\cite{chhieb2024time,chhieb2025fractional,el2024entanglement}. Therefore, the fractional parameter $\tau$ can be regarded as an experimental tuning knob, enabling either fast entanglement generation or longer preservation of correlations. This tunability opens perspectives for the design of robust quantum information and communication protocols, where the control of memory effects and decoherence is essential.

\begin{widetext}
	\begin{minipage}{\linewidth}
		\begin{figure}[H]
			\centering
		\subfigure[]{\label{fig2a}\includegraphics[scale=0.40]{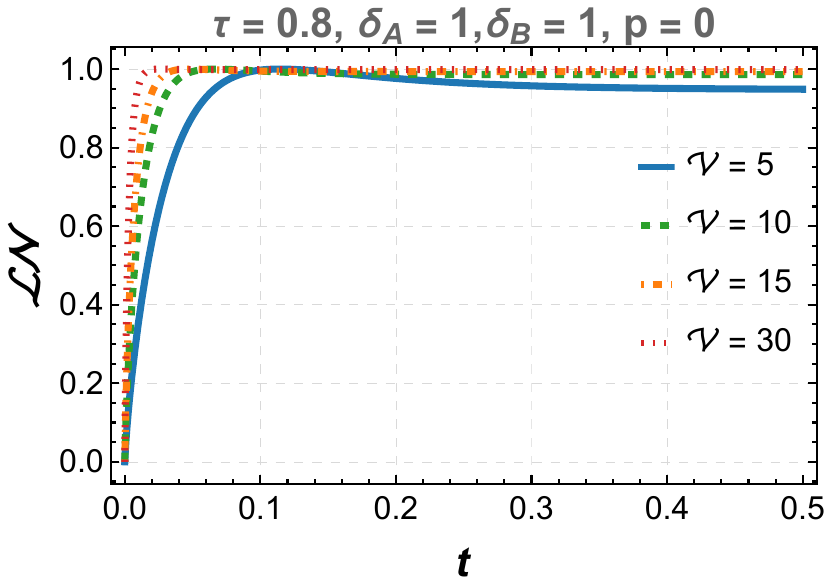}}
			\subfigure[]{\label{fig2b}\includegraphics[scale=0.40]{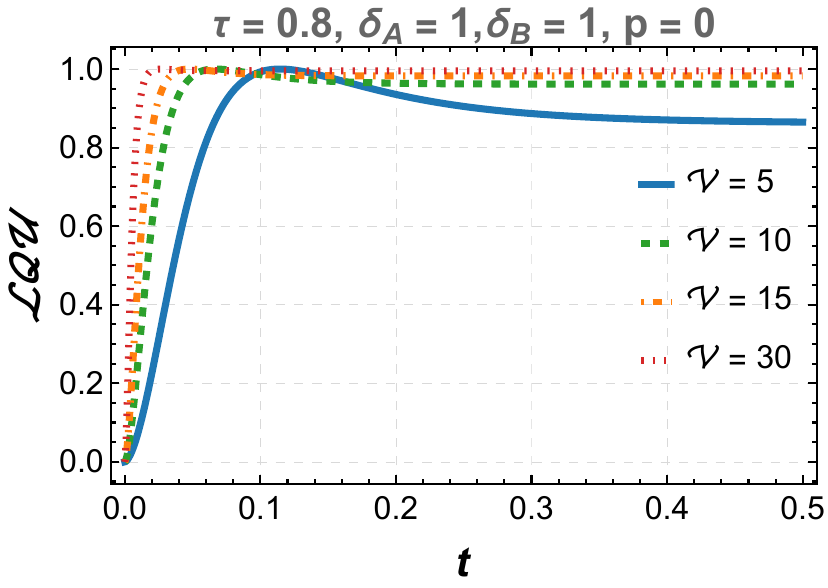}}
			\subfigure[]{\label{fig2c}\includegraphics[scale=0.40]{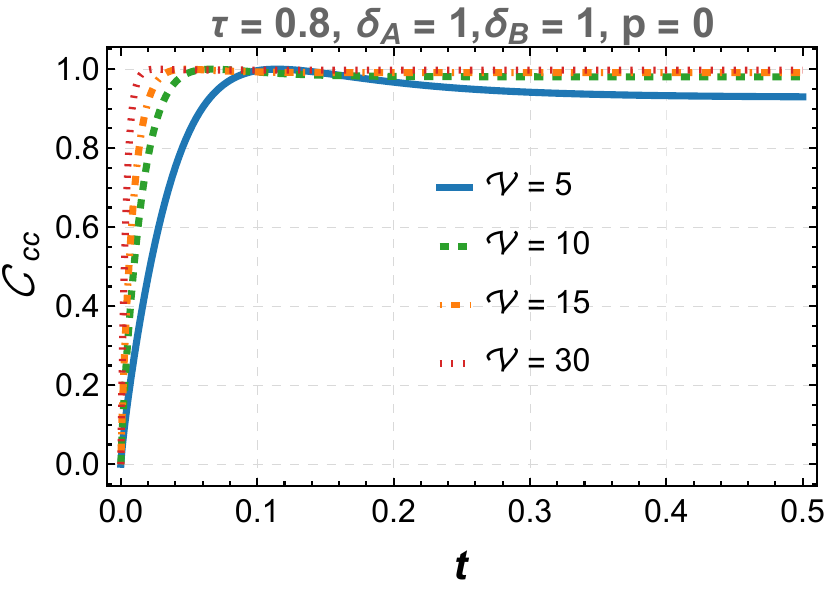}}\\
       \subfigure[]{\label{fig2d}\includegraphics[scale=0.40]{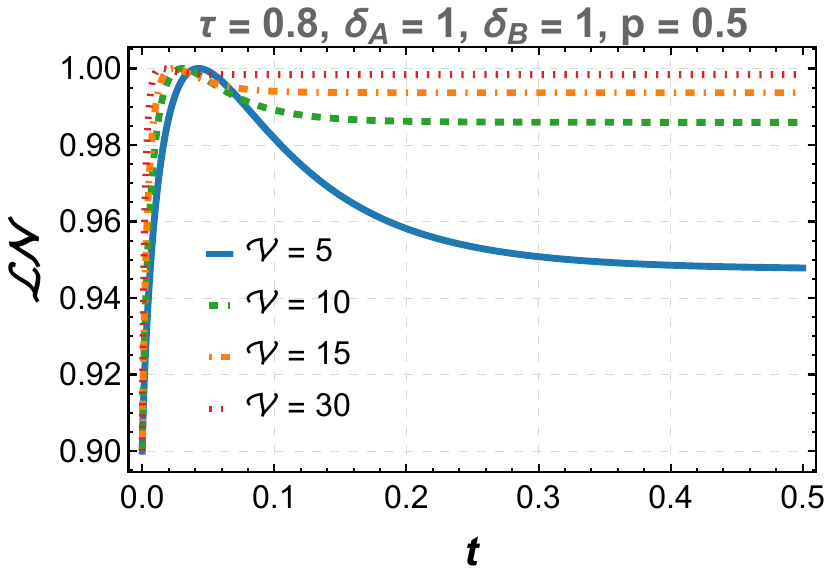}}
			\subfigure[]{\label{fig2e}\includegraphics[scale=0.40]{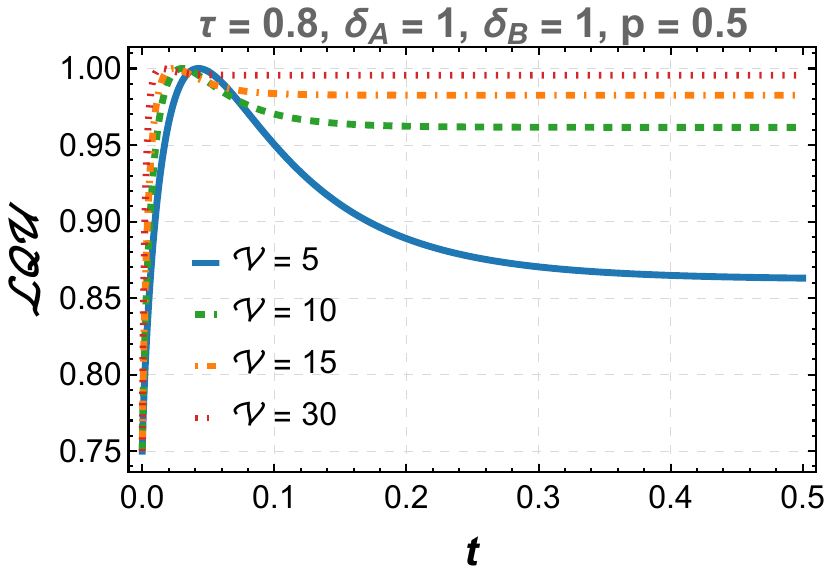}}
			\subfigure[]{\label{fig2f}\includegraphics[scale=0.40]{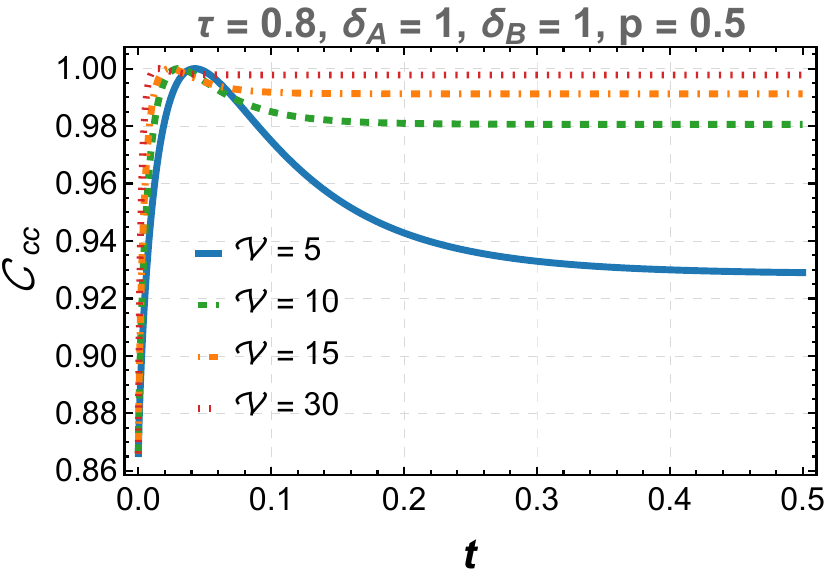}}
\caption{Plots illustrating $\mathcal{LN}$ (\subref{fig2a}-\subref{fig2d}),  
$\mathcal{LQU}$ (\subref{fig2b}-\subref{fig2e}), and  
Correlated coherence (\subref{fig2c}-\subref{fig2f}) as functions of \(t\) for various values of the frequency \(\nu\),  
with \(\tau = 0.8\), \(\delta_A = 1\), \(\delta_B = 1\),  
for \(p=0\) \textit{(top panel)} and \(p=0.5\) \textit{(bottom panel)}.}
\label{figure4}
\end{figure}
\end{minipage}
\end{widetext} 

Figure~\ref{figure4} illustrates the time evolution of quantum correlations in a system of two coupled double quantum dots (DQDs), highlighting the influence of the frequency parameter ($\nu$) on three fundamental quantum resources: logarithmic negativity ($\mathcal{LN}$), local quantum uncertainty ($\mathcal{LQU}$), and correlated coherence ($\mathcal{C}_{cc}$). The analysis is carried out for increasing values of $\nu = 5, 10, 15, 30$, with a fixed fractional parameter $\tau = 0.8$ and tunneling amplitudes $\delta_A = \delta_B = 1$. For the separable state $|11\rangle$ ($p = 0$) (Fig.~\ref{fig2a}, \ref{fig2b} and \ref{fig2c}), $\mathcal{LN}$ increases rapidly from zero to reach a maximum value near 1, which gives a perfect entanglement generation. The increase is faster for larger values of $\nu$. For example, for $\nu = 30$, the maximum entanglement appears at $t \approx 0.05$. Conversely, for $\nu = 5$, the same level is reached around $t \approx 0.15$. This acceleration with increasing $\nu$ explains the more pronounced energy exchange, facilitating the generalization and rapid stabilization of quantum correlations~\cite{elghaayda2022local,fanchini2010entanglement}.
The same behavior for Local Quantum Uncertainty (LQU), for a frequency higher than 30, it quickly reaches its maximum value, indicating strong non-classical correlations, resulting from the moderate fractional dynamics considered ($\tau = 0.8$).
The correlated coherence ($\mathcal{C}_{cc}$) increases strongly and expects maximum values close to 1 for higher values of frequency $\nu = 30$. This increase indicates that the system for higher frequencies not only promotes entanglement, but also preserves quantum coherence in the first moments by reducing the initial decoherence.~\cite{hichri2004entangled}.
For the partially entangled initial state ($p = 0.5$) (Figs.~\ref{fig2d}, \ref{fig2e} and \ref{fig2f}), the three quantum resources start with high values ($\mathcal{LN}(0) \approx 0.90 , \mathcal{LQU}(0) \approx 0.75 and \mathcal{C}_{cc}(0) \approx 0.86$), showing strong initial correlations. A small increase in a very short time (reaching values close to 1), accompanied by a successive decay depending on $\nu$. Despite this decrease, the three measurements $\mathcal{LN}(t)$, $\mathcal{LQU}(t)$ and $\mathcal{C}_{cc}(t)$ remain constant regarding the frequency effect $\nu$ if the same thing for the separable state, showing a more marked dissipation induced by a less powerful energy exchange and an increased sensitivity to decoherence~\cite{chaouki2022dynamics}.
Finally, we concluded that increasing the frequency $\nu$ improves the stability of quantum resources. Higher values encourage rapid generation of correlations and decrease energy losses, making frequency tuning an essential parameter to improve the preparation of quantum information protocols in a Quantum Dot system.~\cite{filgueiras2020thermal,you2011atomic}.\\

\begin{widetext}
	\begin{minipage}{\linewidth}
		\begin{figure}[H]
			\centering
		\subfigure[]{\label{fig3a}\includegraphics[scale=0.40]{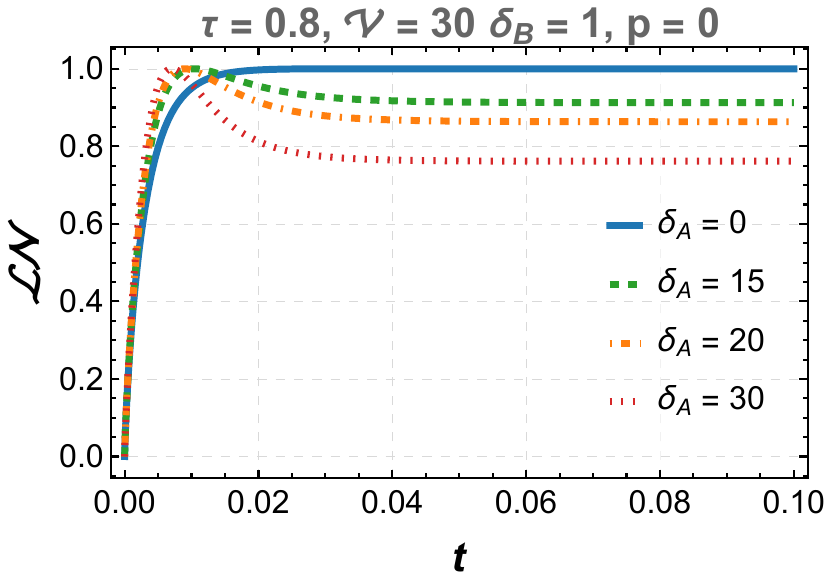}}
			\subfigure[]{\label{fig3b}\includegraphics[scale=0.40]{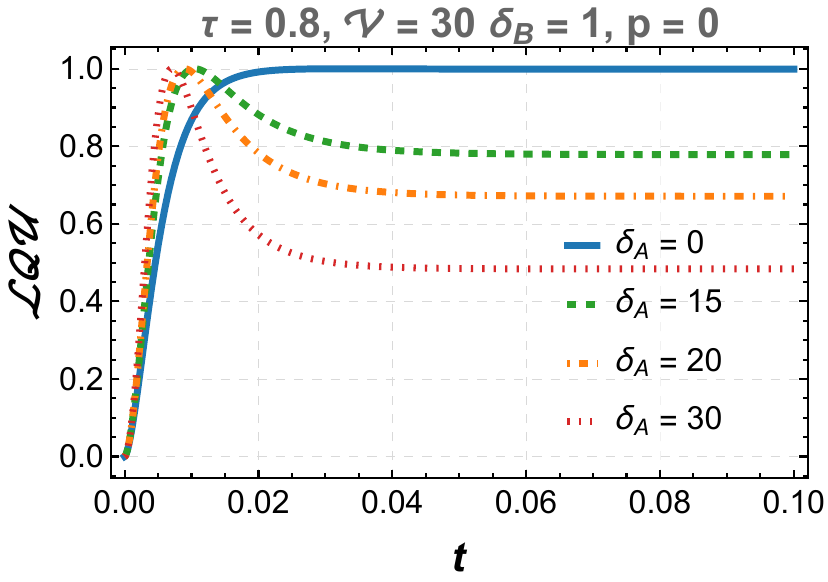}}
			\subfigure[]{\label{fig3c}\includegraphics[scale=0.40]{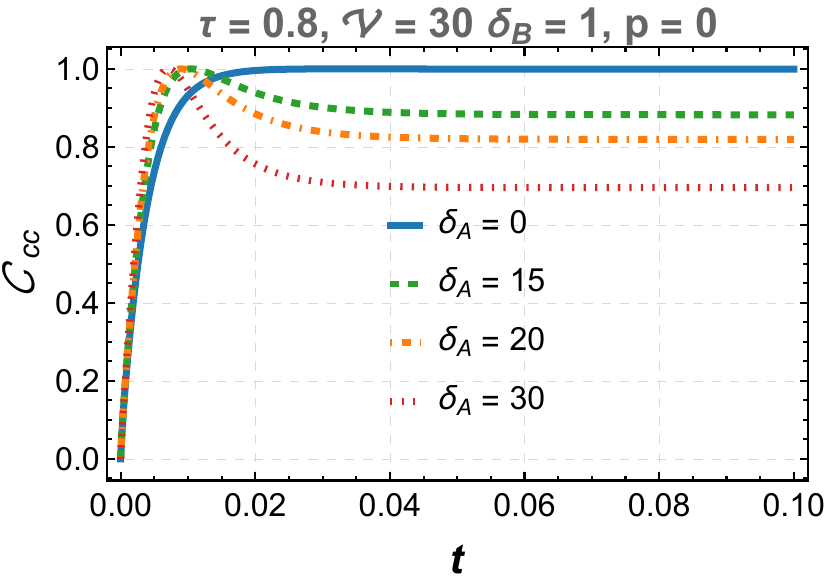}}\\
       \subfigure[]{\label{fig3d}\includegraphics[scale=0.40]{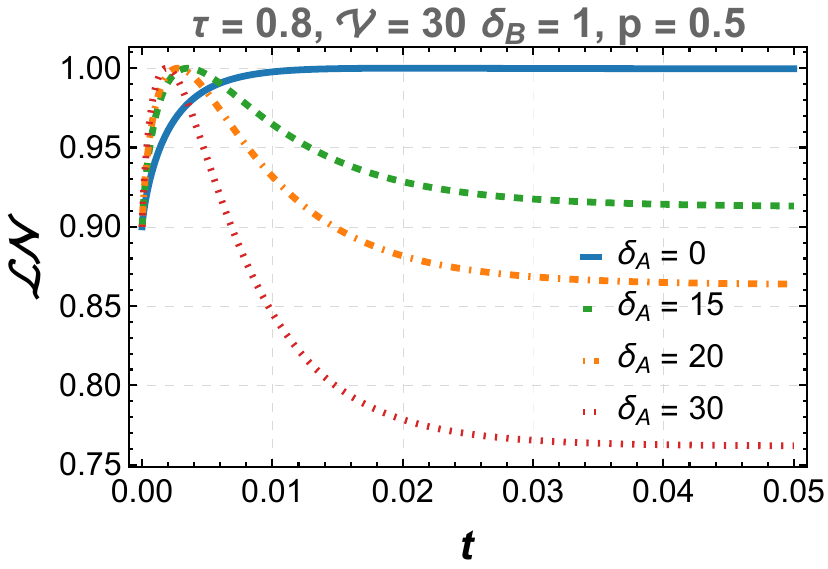}}
			\subfigure[]{\label{fig3e}\includegraphics[scale=0.40]{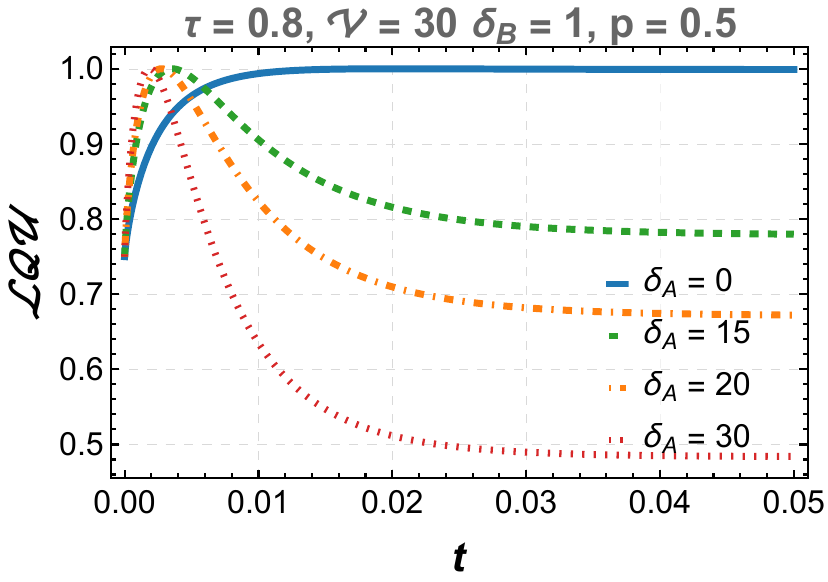}}
			\subfigure[]{\label{fig3f}\includegraphics[scale=0.40]{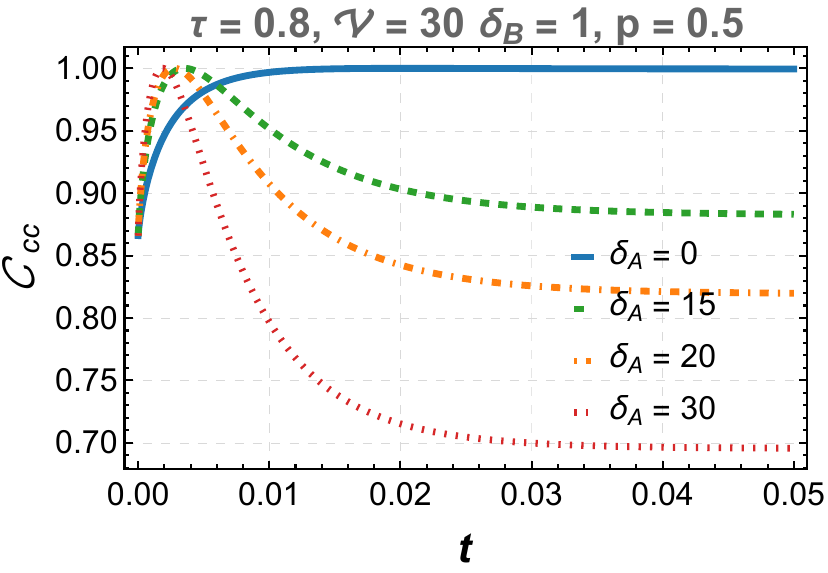}}
\caption{Plots illustrating $\mathcal{LN}$ (\subref{fig3a}-\subref{fig3d}),  $\mathcal{LQU}$ (\subref{fig3b}-\subref{fig3e}), and  Correlated coherence (\subref{fig3c}-\subref{fig3f}) as functions of \(t\) for various values of  the tunneling
 coupling strength \(\delta_A\), with \(\tau = 0.8\),\(\mathcal{V} = 30\) , \(\delta_B = 1\), for \(p=0\) \textit{(top panel)} and \(p=0.5\) \textit{(bottom panel)}.}
\label{figure5}
\end{figure}
\end{minipage}
\end{widetext}
Figure \ref{figure5} illustrates the influence of the tunneling coupling strength (\(\delta_A\)) on the temporal behavior of key quantum resources—logarithmic negativity (\(\mathcal{LN}\)), local quantum uncertainty (\(\mathcal{LQU}\)), and correlated coherence (\(\mathcal{C}_{cc}\))—in a system of two coupled double quantum dots subject to fractional Caputo dynamics with \(\tau = 0.8\). For the initially separable state (\(p = 0\)), quantum correlations dynamically emerge and quickly reach high values when \(\delta_A = 30\). However, the time required to attain these maximal values decreases significantly as \(\delta_A\) decreases. For instance, \(\mathcal{LN}_{\text{max}} \approx 1\) is observed at \(t \approx 0.05\) for \(\delta_A = 30\), whereas it already appears at \(t \approx 0.01\) for smaller values of \(\delta_A\). In the case of a partially entangled initial state (\(p = 0.5\)), the quantum resources start with relatively high values, followed by a pronounced peak under the effect of \(\delta_A\), but they decay more rapidly as \(\delta_A\) increases.  
These findings demonstrate that even under moderately non-Markovian dynamics, strong tunneling coupling asymmetry introduces internal dissipation that weakens entanglement, nonclassical correlations, and coherence. This behavior is consistent with previous results reported in \cite{afsaneh2020robust}.

Figure \ref{figure6} represents the time evolution of three key quantum resources — logarithmic negativity ($\mathcal{LN}$), local quantum uncertainty ($\mathcal{LQU}$) and correlated coherence ($\mathcal{C}_{cc}$) — under different values of the tunneling amplitude $\delta_B$, in a fractional quantum dynamics framework characterized by $\tau = 0.8$. with the other parameters are fixed at $V = 30$ and $\delta_A = 1$. We considered two initial states: ($p = 0$) a separable state and ($p = 0.5$) a partially entangled state, which gives a relevant comparison of the impact of the initial state on entanglement and tunneling asymmetry with the quantum characteristics of the system. In both scenarios, the three resources initially increase due to the coherent dynamics induced by inter-dot interaction. However, the long-term behavior strongly depends on the tunneling strength $\delta_B$. For small values ($\delta_B = 0.15$), quantum correlations—including entanglement, non-classical correlations, and coherence—are preserved over a longer time.On the other hand, for large $\delta_B$ (e.g., $\delta_B = 30$), all resources are evidently degraded, but the big values are still there, highlighting the adverse effect of asymmetric tunneling on the system's quantum characteristics, regardless of the initial state.
Although the partially entangled state ($p = 0.5$) leads to higher initial values of $\mathcal{LQU}$, $\mathcal{LN}$, and $\mathcal{C}_{cc}$, these advantages are not robust in the presence of strong tunneling. Indeed, initially enhanced resources degrade even faster, while the separable state ($p = 0$) does not benefit from any reinforcement under strong asymmetry. This indicates that a higher degree of initial entanglement does not guarantee better preservation of quantum correlations in the long term when the system is subject to strong dissipative coupling.
These findings are consistent with recent studies on non-Markovian quantum dynamics and the fragility of quantum resources in asymmetrically coupled systems~\cite{li2018concepts, breuer2016colloquium}.
\begin{widetext}
	\begin{minipage}{\linewidth}
		\begin{figure}[H]
			\centering
		\subfigure[]{\label{fig4a}\includegraphics[scale=0.40]{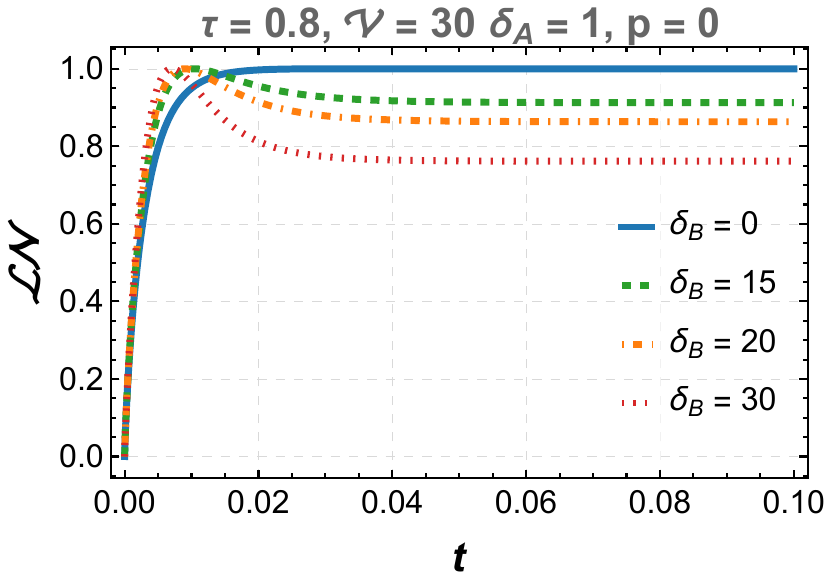}}
			\subfigure[]{\label{fig4b}\includegraphics[scale=0.40]{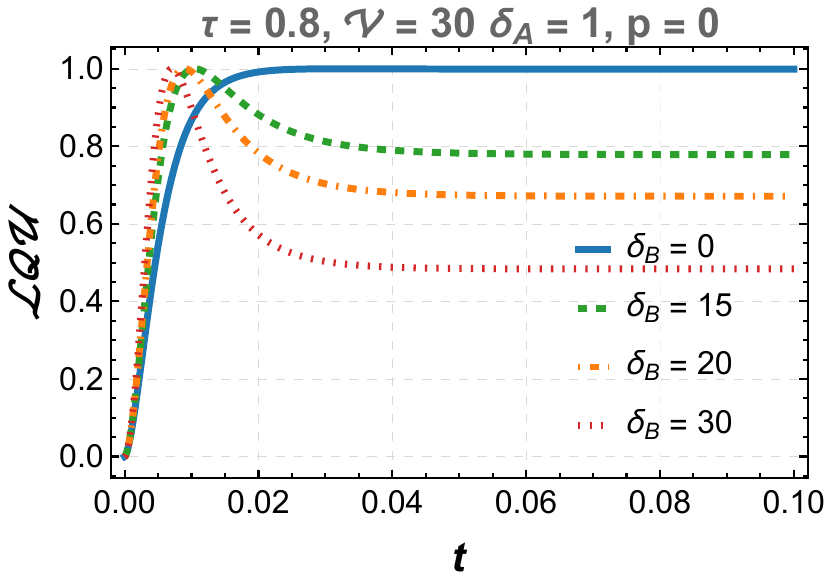}}
			\subfigure[]{\label{fig4c}\includegraphics[scale=0.40]{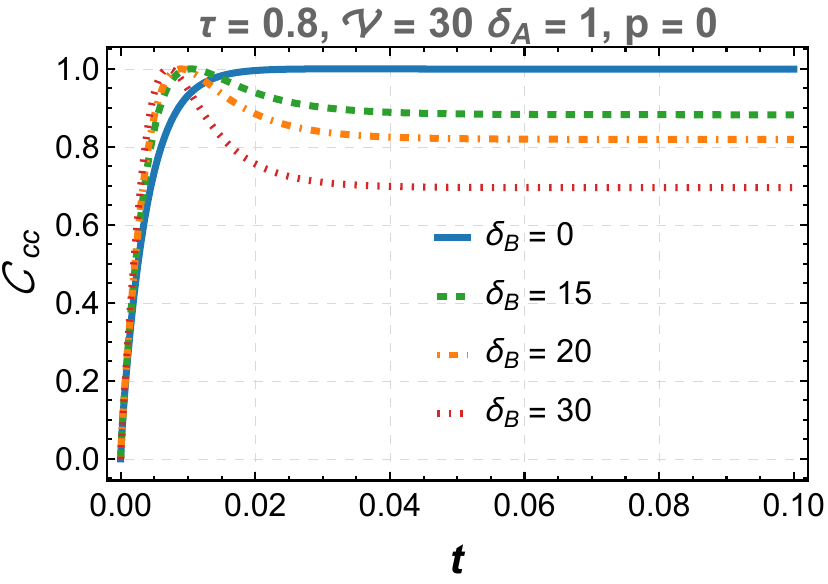}}\\
       \subfigure[]{\label{fig4d}\includegraphics[scale=0.40]{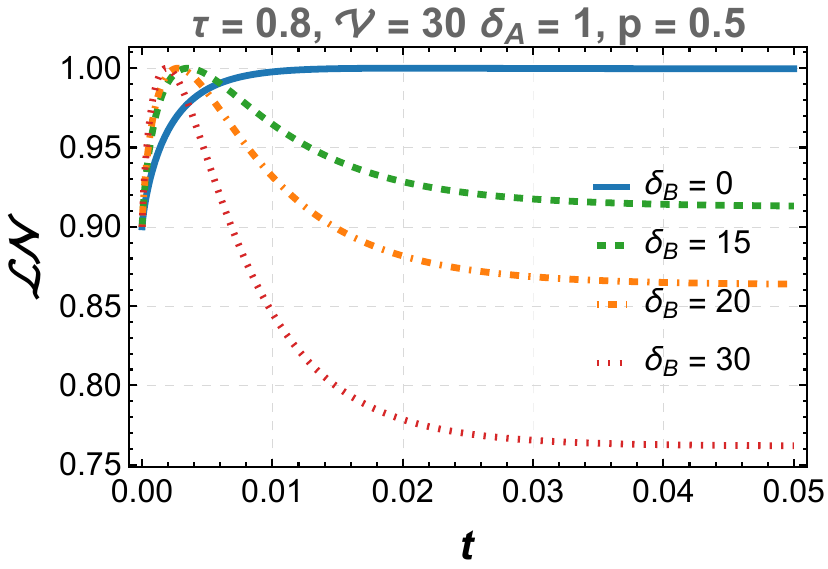}}
			\subfigure[]{\label{fig4e}\includegraphics[scale=0.40]{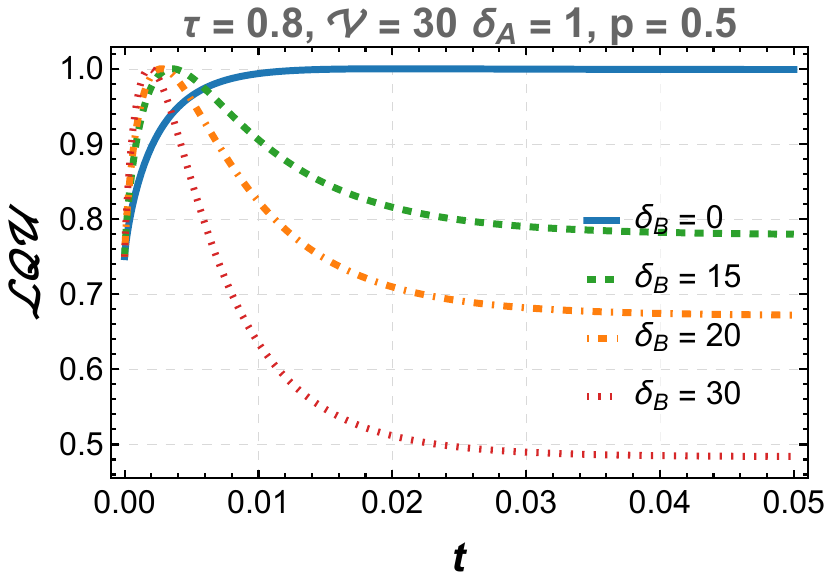}}
			\subfigure[]{\label{fig4f}\includegraphics[scale=0.40]{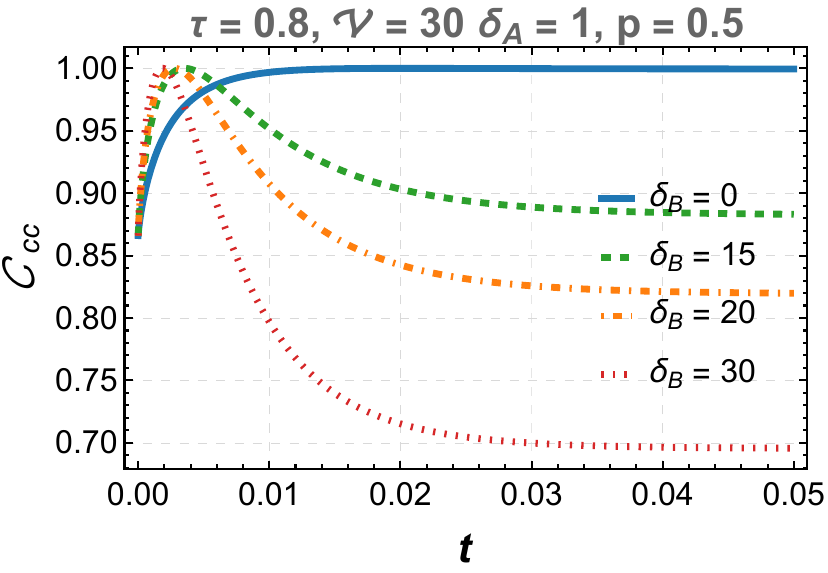}}
\caption{Plots illustrating $\mathcal{LN}$ (\subref{fig4a}-\subref{fig4d}),  $\mathcal{LQU}$ (\subref{fig4b}-\subref{fig4e}), and  Correlated coherence (\subref{fig4c}-\subref{fig4f}) as functions of \(t\) for various values of  the tunneling
 coupling strength \(\delta_A\), with \(\tau = 0.8\),\(\mathcal{V} = 30\) , \(\delta_B = 1\), for \(p=0\) \textit{(top panel)} and \(p=0.5\) \textit{(bottom panel)}.}
\label{figure6}
\end{figure}
\end{minipage}
\end{widetext}

\section{Conclusion}
\label{sec5}

In our work, we studied the temporal evolution of entanglement (via logarithmic negativity), local quantum uncertainty, and correlated coherence in a system of two coupled double quantum dots (DQDs) governed by a fractional Schrödinger equation. Results show that fractional dynamics ( $\tau$ $<$ 1 ) introduces memory effects that lead to non-monotonic behavior, delayed decoherence, and even instability of quantum correlations.
Weakly entangled or separable initial states can dynamically generate quantum correlations in symmetric configurations with strong coupling asymmetry. Tunnel coupling tends to accelerate entanglement and coherence, even under moderate non-Markovian effects.
The obtained results are consistent with recent studies on fractional dynamics to describe non-Markovian behavior in open quantum systems, which emphasize the slowdown of decoherence induced by fractional equations. Similar findings were reported in the study~\cite{chhieb2024metrological, chhieb2025fractional, el2022dynamics}, where the use of fractional methods highlights the interest of fractional quantum dynamics as a versatile framework for the improvement and potential stabilization of quantum resources beyond standard Markovian approximations. Several promising research directions emerge from this work: extending the model to include thermal effects and realistic open environments to better capture temperature-induced decoherence; exploring multi-partite and higher-dimensional quantum systems, such as hybrid qubit-qubit platforms, to reveal collective and critical memory effects; and experimental validation of predictions using reservoirs designed in solid-state devices or photonic simulators.
Finally, the exploration of the impact of fractional dynamics on quantum resources can open new avenues to the development of non-Markovian quantum sensing protocols. Collectively, these avenues constitute an expanded vision of fault-stable quantum technologies that leverage inherent memory effects to counteract decoherence and maintain quantum predominance. These advances are strongly linked to current proposals for networked quantum services and the formation of the quantum Internet~\cite{networked2025, advances2022, survey2017}.

\section*{Appendix A}
Fractional calculus has emerged as a powerful mathematical tool in quantum physics, extending conventional differential operators to non-integer orders. The Riemann-Liouville integral serves as a foundational element in this framework, naturally leading to fractional evolution equations like the fractional Schrödinger equation (Eq.~\ref{FTSE}).The Riemann-Liouville integral of order $\tau > 0$ for a function $F(t)$ is defined as
\begin{equation}
\mathcal{J}^{\tau} F(t) = \frac{1}{\Gamma(\tau)} \int_{0}^{t} (t - s)^{\tau - 1} F(s)\, ds, \qquad t > 0,
\label{eq:RL_integral_new}
\end{equation}
where $\Gamma(\tau)$ is the Gamma function. This integral operator satisfies the composition properties
\begin{equation}
\mathcal{J}^{\tau}\mathcal{J}^{\beta} F(t) = \mathcal{J}^{\tau+\beta} F(t),
\mathcal{J}^{\tau} t^{m} = \frac{\Gamma(m+1)}{\Gamma(\tau+m+1)} t^{\tau+m}.
\end{equation}

Within the Caputo framework, this integral subsequently defines the fractional derivative of order $\tau$: for a function $F(t)$, it is given by
\begin{equation}
\mathcal{D}^{\tau} F(t) = \mathcal{J}^{n-\tau} \left( \frac{d^{n}}{dt^{n}} F(t) \right)
= \frac{1}{\Gamma(n-\tau)} \int_{0}^{t} \frac{F^{(n)}(s)}{(t-s)^{\tau-n+1}} \, ds,
\end{equation} where $n = \lceil \tau \rceil$. This formula explicitly relates fractional derivatives to integer-order derivatives. A general linear fractional system of order $\tau >0$ then obeys
\begin{equation}
\label{eq:frac_system}
\frac{d^{\tau}}{dt^{\tau}} \mathbf{Y}(t) = \mathbf{A} \mathbf{Y}(t), \qquad \mathbf{Y}(0) = \mathbf{Y}_{0},
\end{equation}
where $\mathbf{Y} \in \mathbb{R}^{n}$, $\mathbf{A} \in \mathbb{R}^{n \times n}$, and the derivatives are taken in the Caputo sense.
The solution of Eq.~(\ref{eq:frac_system}) involves the one-parameter Mittag-Leffler function
\begin{equation}
E_{\tau}(z) = \sum_{k=0}^{\infty} \frac{z^{k}}{\Gamma(\tau k + 1)} .
\end{equation}
A particular solution takes the form
\begin{equation}
\mathbf{Y}(t) = \mathbf{v} \, E_{\tau}(\lambda t^{\tau}),
\end{equation}
where $\mathbf{v}$ is a constant vector and $\lambda$ is related to the eigenvalues of $\mathbf{A}$. Substitution into Eq.~(\ref{eq:frac_system}) yields the eigenvalue equation
\begin{equation}
\mathbf{A} \mathbf{v} = \lambda \mathbf{v} \quad \Longleftrightarrow \quad (\mathbf{A} - \lambda \mathbf{I}_{n}) \mathbf{v} = 0,
\end{equation}
demonstrating that $\lambda$ and $\mathbf{v}$ are spectral components of $\mathbf{A}$. Consequently, the general solution is given by
\begin{equation}
\mathbf{Y}(t) = \sum_{j=1}^{n} c_{j} \, E_{\tau}(\lambda_{j} t^{\tau}) \mathbf{v}_{j},
\end{equation}
with coefficients $c_{j}$ determined by the initial condition $\mathbf{Y}(0)$.

\section*{Declarations}

\subsection*{Funding and/or Conflicts of Interest}
This research received no external funding. The authors declare that they have no conflicts of interest.

\bibliography{references}
\bibliographystyle{unsrt}

\end{document}